\documentstyle[preprint,aps]{revtex}
\begin{document}
\preprint{KAIST-TH 14/96\hspace{.3cm} CBNU-TH 961003 \hspace{.3cm}
          hep-ph/9610504}

\title{\bf The $\mu$ problem, and $B$ and $L$ 
Conservation 
with a Discrete Gauge R Symmetry}

\author{Kiwoon Choi${}^{\dagger}$ , Eung Jin Chun${}^*$, and 
Hyungdo Kim${}^{\dagger}$}

\address{Department of Physics, Korea Advanced Institute of Science
and Technology, Taejon 305-701, Korea${}^{\dagger}$}


\address{Department of Physics, Chungbuk National
University, Cheongju 360-763, Korea${}^*$}



\maketitle

\begin{abstract}

We examine in a generic context how the $\mu$ problem 
can be resolved by means of a spontaneously
broken gauge symmetry.
We then focus on the new scheme 
based on a discrete gauge $R$ symmetry which is
spontaneously broken by nonperturbative hidden sector dynamics triggering
supersymmetry breaking also.
The possibility to suppress  the dangerous baryon and/or lepton
number violating interactions by means of
this discrete
$R$ symmetry is examined also together with
some phenomenological consequences.

\end{abstract}
\def\be{\begin{equation}}
\def\ee{\end{equation}}
\def\bea{\begin{eqnarray}}
\def\eea{\end{eqnarray}}

Despite of its great phenomenological success, the minimal standard
model of particle physics is not considered as a final story,
but as a low energy limit of some extended version.
Perhaps  the most promising extension of the standard model
would be to introduce supersymmetry (SUSY) at the weak scale \cite{nilles}.
One  puzzle in SUSY models is  why 
the Higgs $\mu$ parameter 
is of order  of the weak scale \cite{kim1},
not the order of
the Planck scale $M_P=1/\sqrt{8\pi G_N}$.
Another puzzle  is  
why the  baryon ($B$)   and/or lepton ($L$) number
violating interactions  are  so tiny. 
These puzzles  strongly indicate that
SUSY models should be restricted by some additional
symmetry
other than SUSY and $SU(3)\times SU(2)\times
U(1)$ gauge symmetry.

In regard to the $B$ and $L$ 
conservation,  one usually introduces  
$R$ parity  which forbids  renormalizable
$B$  and/or $L$ violating interactions.
However
$R$ parity allows  $d=5$ operators which would lead
to a too rapid proton decay unless their   coefficients
are smaller than the natural values by  
eight    
orders of magnitude \cite{pdecay}.
This suggests that $B$ and $L$
conservation in SUSY models
may {\it not} be a consequence of $R$ parity, but of
a {\it different symmetry} \cite{pati}.

About the problem of the weak scale $\mu$,
a fully successful solution is expected to explain not only
why $\mu\ll M_P$ but also why the two apparently {\it independent}
mass parameters $\mu$ and $m_{3/2}$ have the same order of magnitude.
(Throughout this paper, we assume that SUSY is broken by a hidden sector 
yielding the {\it weak scale} gravitino mass $m_{3/2}$ \cite{nilles}.)
One popular solution to the $\mu$ problem  is
to assume that the $\mu$ term originates from the K\"{a}hler potential
in the underlying supergravity (SUGRA) model \cite{gm}.
This K\"{a}hler potential 
contribution to $\mu$ 
is automatically of order $m_{3/2}$, however one still needs
a rationale justifying the absence of other contributions 
much larger than $m_{3/2}$.
As was stressed recently \cite{kim2}, such a rationale can be
provided by some
symmetry $X$ which is broken in  appropriate manner.

If this $X$ is a gauge symmetry, which is desirable 
in view of naturalness, its breaking must be a spontaneous one.
Then the low energy lagrangian below the scale
of $X$ breaking is constrained by the selection
rule  defined by the $X$ charges of both the low energy fields
and the symmetry breaking order parameters which are determined by
symmetry breaking vacuum expectation values (VEVs).
It would  be interesting if the $X$ selection rule leads to
the enough suppression of $B$ and/or $L$ violating interactions
as well as the desired relation $\mu\simeq m_{3/2}$.
In this paper, we first examine in a generic context  how 
the relation $\mu\simeq m_{3/2}$ can  arise 
as  a consequence of some spontaneously
broken gauge symmetry. 
We then focus on the 
new scheme based on
a discrete $R$ gauge symmetry ($R_X$), and
examine the possibility that the $R_X$ selection rule leads to 
the enough suppression of $B$ and/or $L$
violations.
There are several attractive features in this scheme.
First of all, $R_X$ is spontaneously broken by  nonperturbative
hidden sector dynamics  triggering SUSY breaking also, and
as a result
all $R_X$ breaking VEVs 
can be described  by a single naturally small   
order parameter $\epsilon\simeq e^{-C/g^2}$ ($C=O(\pi^2)$)
which is intimately related to 
SUSY breaking as $m_{3/2}\simeq \epsilon^kM_P$
where $k$ is a positive integer.
Then the $R_X$ selection rule is  simple enough
to be easily applied for low energy physics.
Also since $R_X$ is spontaneously broken, some of $B$ and/or $L$ violating
coefficients are not vanishing and can be close
to the current experimental limits,
leading to
interesting phenomenologies  which can be tested in near future.
Finally this scheme fits well with string theory
since a discrete
gauge $R$ symmetry appears often in compactified string theories
together with a  hidden sector 
breaking both SUSY and $R_X$ \cite{banks1}.

To be concrete,
let us
consider the most general effective superpotential
in the minimal supersymmetric standard model (MSSM):
\bea \label{Weff}
&&W_{\rm eff}=
(\lambda_UH_uQU^c+\lambda_DH_dQD^c+\lambda_EH_dLE^c)\nonumber \\
&&+(\mu H_uH_d+\mu^{\prime}H_uL)
+(\lambda LLE^c +\lambda^{\prime}LQD^c+\lambda^{\prime\prime}U^cD^cD^c) 
\nonumber \\
&&+\frac{1}{M_P}(\lambda_1QQQL+\lambda_2U^cU^cD^cE^c+\lambda_3QQQH_d+...),
\eea 
where $H_u$ and $H_d$ denote the two Higgs doublets, $Q$ and $L$
are the quark and lepton doublets,  $U^c$, $D^c$ and $E^c$
are the quark and lepton singlets, and 
the ellipsis stand for $d\geq 5$ terms other than
the two explicitly written $d=5$ terms.
Here the generation indices of the quark and lepton
superfields are omitted. 
For a successful electroweak symmetry breaking,
one needs
$\mu\simeq 10^{-16} M_P$.
One also needs 
$\mu^{\prime}\leq 10^{-21} M_P$
in order that  the resulting neutrino masses are cosmologically
safe \cite{hs}. 
Furthermore,  proton stability  provides the following constraints
\cite{pdecay}:
$\lambda^{\prime}\lambda^{\prime\prime}\leq 10^{-26}$,
$\lambda^{\prime}\lambda_3\leq 10^{-10}$,
$\lambda_1\leq 10^{-8}$, and $\lambda_2\leq 10^{-8}$
when one assumes superparticle masses of order $10^2$ GeV.
The above constraints on $\mu^{\prime}$ and $\lambda_2$
have been derived  for a generic form of
soft SUSY breaking
and thus  can be relaxed by 
several orders
of magnitude  if soft terms satisfy certain 
universality conditions \cite{white}.
Also the consideration of $n$-$\bar{n}$ oscillation would provide
a constraint on
$\lambda^{\prime\prime}$, however recent studies show that
it is not so significant \cite{gs}.
(Here we ignore the flavor structure of
these constraints since we will consider the suppression
of couplings by means of
a generation-independent symmetry.)
Note that
$R$ parity allows $\lambda_1$ and $\lambda_2$ to be of order unity,
and thus  {\it not} guarantee the enough  proton stability.

To proceed, 
let us imagine a spontaneously broken
gauge symmetry $X$.
To be useful for low energy physics,
it is desirable that $X$ breaking
is described by
a single order parameter $\epsilon$. 
This would be the case if  $X$-breaking
is due to  the VEVs of a set of holomorphic
operators $\{\Gamma_i\}$ (of course with nonzero $X$ charges $\{q(\Gamma_i)\}$)
obeying
\be \label{gamma}
\langle\Gamma_i \rangle \,  \simeq \, \theta (X_i)\epsilon^{X_i}
M_P^{d_i}
\ee
where $X_i=q({\Gamma_i})/q({\epsilon})$, 
$d_i$ denotes the dimension of $\Gamma_i$,
and $\theta (x)=1$ for {\it non-negative integer} $x$,  
while $\theta (x)=0$ for other values of $x$.

How can the relation $\mu\simeq m_{3/2}$
arise as a consequence of the $X$ selection rule?
Generic selection rule defined by $\epsilon$
would give rise to
\be
\mu\simeq \theta(l_1)M_P\epsilon^{l_1}+m_{3/2}[\theta(l_2)\epsilon^{l_2}
+\theta(l_3)\bar{\epsilon}^{l_3}],
\ee
where  $l_i$'s   
satisfy
\bea 
&&l_1q(\epsilon)+q(H_uH_d)=q(\theta^2),
\nonumber \\
&&-l_2q(\epsilon)=
l_3q(\epsilon)=q(H_uH_d).
\eea
If $X$ is a discrete symmetry, say $Z_N$, these relations between the $X$
charges  are defined
modulo $N$.
Here $q(\theta^2)$ denotes the $X$ charge of the Grassmann variable $\theta^2$
and also of
the gravitino mass $m_{3/2}=e^{K/2M_P^2}W/M_P^2$.
Note that 
the first part of the above $\mu$  originates from the
SUGRA
superpotential, 
while the second  part proportional to ${m}_{3/2}$ is from
the SUGRA K\"{a}hler potential.

Let us first consider the case of $q(\theta^2)=0$.
Obviously then $l_1=l_2$. Furthermore,
if $l_3=0$, then $l_1=l_2=0$ also, yielding $\mu\simeq M_P$.
Thus neither $l_2$ nor $l_3$  can be zero.
Then there are essentially two ways to get $\mu\simeq m_{3/2}$. 
In the first case of $\epsilon\ll 1$, one can ignore
$m_{3/2}\epsilon^{l_2}$ and $m_{3/2}\bar{\epsilon}^{l_3}$, and
thus $\mu\simeq \theta(l_1)M_P\epsilon^{l_1}$.
In order for this $\mu$ to be of order $m_{3/2}$,
the size of $\epsilon$ 
should  be correlated with the size of $m_{3/2}$
as $|\epsilon|^{l_1}\simeq
|m_{3/2}|/M_P$ for some positive integer $l_1$.
This indicates that the $X$ breaking scale should be determined
by the size of the $X$-invariant quantity $m_{3/2}$.
In fact, this way to solve the $\mu$ problem
has been considered before by several authors \cite{yanagi}.
As a simple example, one may consider a model with a (quasi-flat) field
$\phi$ whose superpotential contains 
$\phi^{N}/M_P^{N-3}$ 
and $\phi^{N-2}H_uH_d/M_P^{N-3}$.
The model 
is invariant under $X=Z_N$ with $q(\phi)=1$, $q(H_uH_d)=2$
and so on. 
The scalar potential of $\phi$  
contains the soft breaking terms 
$m_0^2|\phi|^2$ and 
$(A_0{\phi^{N}}/{M_P^{N-3}}+{\rm h.c})$ as well
as the supersymmetric part 
$N^2{|\phi|^{2(N-1)}}/{M_P^{2(N-3)}}$,
where $m_0$ and $A_0$ are 
of order $m_{3/2}$.  
Minimizing this potential 
for $|A_0|^2> N^2m_0^2$, 
one finds 
$|\epsilon|=|\langle \phi\rangle|/M_P\simeq
(|m_{3/2}|/M_P)^{1/(N-2)}$, yielding 
$|\mu|\simeq |\epsilon|^{N-2}M_P\simeq |m_{3/2}|$ as desired. 
The second way to get $\mu\simeq m_{3/2}$ 
is to have $\epsilon$ which is somewhat close
to the unity. In this case, if $l_3=1$ and also the Eq.~(4) does not
allow any non-negative integer $l_1$, 
we would have $\mu\simeq \bar{\epsilon}m_{3/2}$.
(Here we ignore the possibility that $l_1$ is so large
that $\epsilon^{l_1}\simeq m_{3/2}/M_P$ even when 
$\epsilon$ is close to the unity. Note that this requires
that some $X$ charges are abnormally large.)
This scheme has been discussed  recently in the context of abelian
horizontal symmetry with
$\epsilon\simeq 0.2$ (Cabibbo angle) \cite{nir}.

Let us now consider the case that
$q(\theta^2)\neq 0$ and thus  $X$ is a gauge $R$ symmetry.
If $X=U(1)_R$, it 
would be  spontaneously
broken at $M_P$ due to the Fayet-Iliopoulos term \cite{dreiner}, which
yields a rather large $\epsilon=\langle\phi\rangle/M_P \simeq 1/10$.
Although  
the $U(1)_R$ selection rule with $\epsilon\simeq 1/10$
can lead $\mu\simeq m_{3/2}$ (or $\mu\simeq |\epsilon| m_{3/2}$)
for an appropriate $U(1)_R$ charge
assignment \cite{chun0}, such scheme has an unattractive
feature that the simplest form of
the $U(1)_R$ selection rule for $m_{3/2}$, i.e.\ the relation 
$m_{3/2}=\epsilon^kM_P$ resulting from Eq.~(2) applied for the hidden sector
superpotential, is {\it not} valid anymore 
unless the positive integer $k=q(\theta^2)/q(\epsilon)$
is abnormally large. 
In this regard, a more interesting possibility
is that $X$ is a discrete gauge $R$ symmetry ($R_X$), allowing
$\epsilon\ll 1/10$.
In fact, the  simple selection rule 
$m_{3/2}=\epsilon^k M_P$
indicates that
local SUSY breaking  and  $R_X$ breaking
have a common dynamical origin.
We are then led to the scenario in which
$R_X$ is broken by nonperturbative hidden sector dynamics
triggering local SUSY breaking also.
In order for the $R_X$ selection rule of Eq.~(3) yield
$\mu\simeq m_{3/2}$, we need $q(H_uH_d)=0$ leading to
$l_1=k$ and $l_2=l_3=0$.

From now on, we focus on the scheme based
on $R_X$ and  discuss first  how nonperturbative hidden sector dynamics
\cite{seiberg} yields   
$R_X$  breaking pattern which is described by a single order parameter
$\epsilon$ as Eq.~(\ref{gamma}).
As a simple example, let us  consider a  hidden sector
with $SU(N_c)$ gauge group
and $N_f$ flavors of hidden quarks $(\Phi+\Phi^c)$ together
with a singlet $A$. The tree level hidden sector superpotential
is given by
\be \label{tree}
W_{\rm hid}=\frac{A^{m+3}}{M_P^m}+\frac{A^{n+1}}{M_P^n}\Phi\Phi^c,
\ee
which is invariant under $R_X=Z_N$ with the $Z_N$ charges satisfying
$(m+3)q(A)=q({\theta^2})$ and $(n+1)q(A)+q({\Phi\Phi^c})=q({\theta^2})$
modulo $N$.
(In this paper,  the integer  $Z_N$ charge $q$ is always defined by the
transformation factor $e^{i2\pi q/N}$.)
Also in $W_{\rm hid}$, we have ignored 
the flavor indices of hidden quarks and also
the dimensionless coefficients which are presumed to be of order unity.
Since $Z_N$ is a  gauge symmetry, it is required to be anomaly free.
This would be
achieved by adopting the Green-Schwarz mechanism
in which the hidden sector gauge kinetic function $f_H$  transforms nonlinearly
under $Z_N$ as 
$f_H\rightarrow f_H+ik_H/4\pi N$,
where $k_H$ is an integer and
$f_H$ is normalized as $\langle f_H\rangle=\frac{1}{g_H^2}+i\frac{\theta_H}{8\pi^2}$
for the hidden sector gauge coupling constant $g_H$ 
and the vacuum angle $\theta_H$ \cite{banks2}.
It is then straightforward  \cite{seiberg} to find that strong $SU(N_c)$ dynamics 
leads to the vacuum configuration  described by
$$\langle A \rangle\simeq \epsilon^{X_A} M_P, 
\langle \Phi\Phi^c\rangle \simeq \epsilon^{X_{\Phi\Phi^c}} M_P^2,
\langle W_H^aW_H^a\rangle \simeq \epsilon^k M_P^3,
$$
where  $X_A=q(A)/q({\epsilon})=1$, $X_{\Phi\Phi^c}=
q({\Phi\Phi^c})/q({\epsilon})=(m-n+2)$,
$k=q({\theta^2})/q({\epsilon})=(m+3)$,
and the symmetry breaking order parameter is given by
$$\epsilon
=e^{-f_H/((m+3)N_c-(n+1)N_f)}.
$$
Here $W^a_H$  denotes the chiral gauge superfield 
for the hidden sector $SU(N_c)$ gauge multiplet.

Obviously  the above symmetry  breaking VEVs 
obey the simple rule of Eq.~(\ref{gamma}).
In fact, 
the value of $k=q(\theta^2)/q(\epsilon)$ is quite relevant for 
phenomenological applications of $R_X$  since 
it determines the size
of $\epsilon$ through the relation $m_{3/2}/M_P=\epsilon^k\simeq
10^{-(15\sim 16)}$.
The above example of SUSY QCD type hidden
sector gives rise to $k=(m+3)\geq 3$.
However the value of $k$  can be smaller in other type
of hidden sector. 
For instance,  hidden sector
without any  singlet $A$ can give $k=2$.
Also for  pure Yang-Mills hidden
sector without any matter fields,
the vacuum configuration 
is described simply by the gaugino condensation
$\langle W_H^aW_H^a\rangle \simeq \epsilon^{k} M_P^3$,
where now $k=q({\theta^2})/q({\epsilon})=1$ for the symmetry
breaking order parameter $\epsilon=e^{-f_H/N_c}$.

One can see more explicitly how  
$R_X$ spontaneously broken by the VEVs obeying the rule of
Eq.~(\ref{gamma})
leads naturally to $\mu\simeq m_{3/2}$ when $q(H_uH_d)=0$.
Generic $R_X$-invariant SUGRA models may generate
the $\mu$ term through its K\"{a}hler potential
$K$, and/or  hidden sector gauge kinetic function $f_H$, and/or the 
superpotential $W$:
$$K\ni Z\,H_uH_d,
\quad
f_H\ni \frac{\Gamma}{M_P^2} H_uH_d,
\quad
W\ni \Gamma_W H_uH_d.
$$
Here $Z$ is a $R_X$-invariant function  of
generic hidden sector fields $\phi$  and their conjugates
$\bar{\phi}$, while $\Gamma$ and $\Gamma_W$ are 
holomorphic functions  with  $q(\Gamma)=0$
and $q(\Gamma_W)=q(\theta^2)$.
(Note that ${\rm dim}(Z)={\rm dim}(\Gamma)=0$ while
${\rm dim}(\Gamma_W)=1$.)
It is then straightforward \cite{gm,chun} to  find
$$\mu=e^{K/2M_P^2} \Gamma_W+
\frac{1}{4M_P^{2}}\Gamma W^a_HW^a_H+ 
m_{3/2}Z
-\bar{F}^{\bar{\phi}}\bar{\partial}_{\bar{\phi}}Z,
$$
where $\bar{F}^{\bar{\phi}}=e^{K/2M_P^2}K^{\phi\bar{\phi}}
(\partial_{\phi}W+W\partial_{\phi}K/M_P^2)$
denotes the auxiliary component of $\bar{\phi}$.
Applying  the rule of Eq.~(\ref{gamma})
for the operators contributing to $\mu$,
which is valid as long as $R_X$ breaking VEVs are induced by 
SUSY breaking  nonperturbative
hidden sector dynamics,
all the contributions to $\mu$ 
from the K\"{a}hler potential,  hidden sector
gauge kinetic function, and the SUGRA superpotential are 
of order $m_{3/2}$.

Discrete gauge $R$ symmetry ($R_X$) appears often in compactified
string theory which generically contains a  hidden sector
for nonperturbative dynamics breaking
both SUSY and $R_X$ \cite{banks1}.
In compactified string theory with $R_X$,
the effective superpotential $W_0$ of the dilaton superfield $S$
and other moduli $M_{\alpha}$ 
would be given by 
$W_0=M_P^3 e^{-f(S, M_{\alpha})}$ where $f$ is a holomorphic function
which  transforms as $f\rightarrow f-i2\pi q(\theta^2)/N$ 
under $R_X$. Note that $R_X$ is nonlinearly realized on the
moduli space of $\{S,M_{\alpha}\}$.
For the hidden sector that we have discussed before,
$f$ corresponds to the hidden sector
gauge kinetic function $f_H$ (multiplied by a model dependent
constant) which is given by
$f_H=S+\delta (M_{\alpha})$ in string perturbation theory.
Here $\delta(M_{\alpha})$ denotes the string
loop threshold correction.
As is well known, we may then encounter the 
dilaton runaway problem.
In this regard,  we note that 
the race-track mechanism \cite{race} for the dilaton stability 
can {\it not} be applied for $W_0\sim e^{-\alpha S}$ ($\alpha={\rm constant}$)
in models with $R_X$,
however the other mechanism relying upon
nonperturbative correction
to the dilaton K\"{a}hler potential 
can work without any difficulty \cite{banks1}.

Let us finally examine the implication of $R_X$  
for the suppression of $B$ and/or $L$ violation.
For simplicity, in this paper we limit ourselves 
to the case that $R_X$ is a generation independent $Z_N$ symmetry.
First of all, we assume that the observed Yukawa couplings
$\lambda_U$, $\lambda_D$, and $\lambda_E$ are {\it not} suppressed
by the small  $\epsilon$ and thus $q(H_uQU^c)=q(H_dQD^c)
=q(H_dLE^c)=q(\theta^2)$. 
Also for $\mu\simeq m_{3/2}$, we require $q(H_uH_d)=0$.
Then 
the $Z_N$ charges of the MSSM fields 
can be determined by $q(\theta^2)$ and the two integers $n_1$ and $n_3$
as $q(Q)=0$, $q(U^c)=-n_1$, $q(D^c)=2q(\theta^2)+n_1$, $q(E^c)=n_1+n_3$, and so on
\cite{iba}.
In our case,  the full definition of $Z_N$ 
includes  the $Z_N$ charge $q(\epsilon)$ for which $k=q(\theta^2)/q(\epsilon)$
is a positive integer, 
and also the integers $k_a$  defining the possible nonlinear transformation
of the observable sector  gauge kinetic functions $f_a$ ($a=3,2,1$ for
$SU(3)_c$, $SU(2)_L$, 
$U(1)_Y$, respectively)
as $f_a\rightarrow f_a+ik_a/4\pi N$.
(Again $f_a=\frac{1}{g_a^2}+i\frac{\theta_a}{8\pi^2}$.)
Note that the values of 
$q(\epsilon)$ and $k$ depend upon what kind of hidden sector
is responsible for $Z_N$ breaking.
Also  a nonzero variation of $f_a$ allows 
the Green-Schwarz  mechanism  canceling the 
$Z_N\times G_a^2$ anomaly for the observable sector
gauge group $G_a$ \cite{banks2}.
Requiring the cancelation 
of the $Z_N\times [SU(3)_c]^2$
and $Z_N\times [SU(2)_L]^2$ mixed anomalies together with $k_2=k_3$
which may be necessary for gauge coupling unification, 
all the $Z_N$ charges  can be determined by three integers:
$k$, $n_1$ and $q(\epsilon)$.

If one writes the low energy effective superpotential 
as
$W_{\rm eff}=\sum_I\lambda_I {\cal O}_I/M_P^{(d_I-3)}$,
where $\{{\cal O}_I\}$ stands for generic $SU(3)_c\times SU(2)_L\times U(1)_Y$
invariant $d_I$-dimensional holomorphic operators,
the $R_X$ selection rule says 
that the dimensionless coefficient $\lambda_I$ 
are  suppressed by the factor
\be \label{lambda}
\theta (A_I) \epsilon^{A_I}
+\frac{{m}_{3/2}}{M_P}[\theta (B_I)\epsilon^{B_I}
+\theta (C_I)\bar{\epsilon}^{C_I}],
\ee
where 
$A_Iq(\epsilon)+q({\cal O}_I)=q(\theta^2)$,
and 
$q({\cal O}_I)=-B_Iq(\epsilon)=C_Iq(\epsilon)$ modulo $N$.
Again, the first part of the above suppression factor is for 
the contribution from the 
SUGRA superpotential, 
while the second part proportional to ${m}_{3/2}$ is for the contribution 
from
the SUGRA K\"{a}hler potential.
In fact, there can be extra piece in $\lambda_I$ arising from
the possible mixing between $L$ and $H_d$ after $R_X$ breaking,
which will be taken  into
account in our analysis.

Within the context described above, we  have examined 
whether the phenomenological constraints 
on $B$ and/or $L$ violation can be fulfilled by means of 
the $R_X=Z_N$ selection rule.
Of course our  approach has  limitation
since some Yukawa couplings, e.g.\ some components
of $\lambda_{U,D,E}$,
are small  without
any suppression by the $Z_N$ selection rule.
However our basic aim here  is to show that 
all of the dangerous $B$ and/or  $L$ violating couplings can be 
suppressed 
by the discrete $R$ symmetry responsible for $\mu\simeq m_{3/2}$.
We thus examine the possibility that
the $Z_N$ selection rule provides  suppression factors {\it smaller}
than  
$10^{-18}$, $10^{-20}$, $10^{-6}$, $10^{-5}$, $10^{-3}$ for 
$\mu^{\prime}/M_P$,
$\lambda^{\prime}\lambda^{\prime\prime}$,
$\lambda^{\prime}\lambda_3$, $\lambda_1$, and $\lambda_2$
respectively.
Note that we can relax the above criterion  by assuming an additional
suppression of the couplings by means of other
symmetry.
A more extensive analysis including a horizontal
$U(1)$ symmetry will be done elsewhere \cite{future}.
At any rate, we found that there are  many 
different $Z_N$-charge assignments (even for 
$N< 10$)  leading to enough suppression of $B$ and $L$
violation: $\{2,7,11,20\}$ different assignments for $N=\{4,6,7,8\}$
respectively. However many of these are {\it not} phenomenologically
distinguishable. In Table 1, we depict 
some charge assignments which lead to
phenomenologically distinguishable
suppression factors for the $B$ and/or $L$ violating
coefficients. The results show that  many charge assignments 
can lead to  
interesting $B$ and/or $L$ violating phenomenologies which are close
to the current experimental limits.

To conclude, we have examined in a generic context
how  the 
$\mu$ problem can be resolved by means of
a spontaneously broken gauge symmetry.
A particular attention was paid for
the new  scheme 
based on a discrete gauge $R$ symmetry  ($R_X$)
which is broken by nonperturbative
hidden sector dynamics triggering SUSY breaking also.
We then examine the possibility that this $R_X$
leads to enough suppression of  the $B$ and/or $L$ 
violating couplings for the case that $R_X$ is a generation-independent
$Z_N$ symmetry.
Even for $N <10$, there are many different $Z_N$ charge assignments
leading to
the enough suppression
of $B$ and/or $L$ violation as well as $\mu\simeq m_{3/2}$.
Models with such a discrete $R$ gauge symmetry can naturally arise
in string theory and some of them
give rise to nonvanishing $B$ and/or $L$ violations
which  may be observed in near
future.


\acknowledgements
This work is supported in part  
by KOSEF Grant 951-0207-002-2 (KC, HK),
Non Directed Research Fund of KRF (EJC).  
EJC is a Brain-Pool fellow.

\begin{table}[ht]
\centering
\caption{ Suppression factors for the dangerous
$B$ and/or $L$ violating couplings due to the $R_X=Z_N$ selection rule.
Here we present only the cases that are phenomenologically distinguishable
from each other. Note that the size of $\epsilon$ is determined
by the relation $\epsilon^k\simeq \mu/M_P\simeq 10^{-16}$.}
\vspace{2mm}
\begin{tabular}{|c|c|c|c||c|c|c|c|c|} 
$~~k~~$ & $q(\epsilon)$ & $n_1$ & $N$
& $\lambda,\lambda^{\prime}$ & $\lambda^{\prime\prime}$
& $\mu^{\prime}/M_P$
& $\lambda_1$
& $\lambda_2$ 
\\ \hline
1 & 3 & 2 & 4 & $\epsilon$ & $\epsilon^3$  
& $\epsilon^2$ & $\epsilon$ & $\epsilon$ 
\\ \hline
2 & 3 & 0 & 7 & $\epsilon^2$ & $\epsilon$ 
 &  $\epsilon^4$ & $\epsilon^6$ & $\epsilon^4$
\\ \hline
3 & 5 & 5 & 6 & $\epsilon^2$ 
& $\epsilon^2$ 
 &  $\epsilon^5$ 
& $\epsilon^3$ & $\epsilon^3$
\\ \hline
3 & 5 & 0 & 6 & $\epsilon$ 
& $\epsilon^3$ 
 &  $\epsilon^4$ 
& $ \epsilon$ & $\epsilon^5$
\\ \hline
3 & 5 & 1 & 6 & $\epsilon^2$ 
& $\epsilon^4$ 
&  $\epsilon^5$ 
& $\epsilon$ & $\epsilon^5$
\\ \hline
3 & 1 & 4 & 6 & $\epsilon$ 
& $\epsilon^5$ 
& $\epsilon^4$ & $\epsilon^5$ & $\epsilon$
\\ \hline
3 & 2 & 2 & 7 & $\epsilon^2$ 
& $\epsilon^4$ 
& $\epsilon^5$ & $\epsilon^2$ & $\epsilon^6$
\\ \hline
3 & 2 & 4 & 7 & $\epsilon$ 
& $\epsilon^3$ 
 & $\epsilon^4$ & $\epsilon^2$ & $\epsilon^6$
\\ \hline
4 & 5 & 1 & 7 & $\epsilon$ 
& $\epsilon^6$ 
 & $\epsilon^5$ & $\epsilon^5$ & $\epsilon$
\\ \hline
4 & 7 & 7 & 8 & $\epsilon^3$ 
& $\epsilon^3$ 
 & $\epsilon^7$ & $\epsilon^4$ & $\epsilon^4$
\\ \hline
5 & 3 & 6 & 8 & $\epsilon$ 
& $\epsilon^7$ 
& $\epsilon^6$ & $\epsilon^5$ & $\epsilon$ 
\end{tabular}
\end{table} 

\end{document}